

\def\today{
  \number\day \space \ifcase\month\or
  January\or February\or March\or April\or May\or June\or
  July\or August\or September\or October\or November\or December\fi
\space\number\year}

\long\def\suppress#1{}

\def\boringfonts{y}  
\input epsf \def\figflag{y}  
\def\answ{b }
\input harvmac.tex


\def\fonttest{y}

\ifx\boringfonts\fonttest
\else

\fi

\hyphenation{anom-aly anom-alies coun-ter-term coun-ter-terms
dif-feo-mor-phism dif-fer-en-tial super-dif-fer-en-tial dif-fer-en-tials
super-dif-fer-en-tials reparam-etrize param-etrize reparam-etriza-tion}


%
%
%
\newwrite\tocfile\global\newcount\tocno\global\tocno=1
\ifx\bigans\answ \def\tocline#1{\hbox to 320pt{\hbox to 45pt{}#1}}
\else\def\tocline#1{\line{#1}}\fi
\def\toclead{\leaders\hbox to 1em{\hss.\hss}\hfill}
\def\tnewsec#1#2{\newsec{#2}\xdef #1{\the\secno}
\ifnum\tocno=1\immediate\openout\tocfile=toc.tmp\fi\global\advance\tocno
by1
{\let\the=0\edef\next{\write\tocfile{\medskip\tocline{\secsym\ #2\toclead\the
\count0}\smallskip}}\next}
}
\def\tnewsubsec#1#2{\subsec{#2}\xdef #1{\the\secno.\the\subsecno}
\ifnum\tocno=1\immediate\openout\tocfile=toc.tmp\fi\global\advance\tocno
by1
{\let\the=0\edef\next{\write\tocfile{\tocline{ \ \secsym\the\subsecno\
#2\toclead\the\count0}}}\next}
}
\def\tappendix#1#2#3{\xdef #1{#2.}\appendix{#2}{#3}
\ifnum\tocno=1\immediate\openout\tocfile=toc.tmp\fi\global\advance\tocno
by1
{\let\the=0\edef\next{\write\tocfile{\tocline{ \ #2.
#3\toclead\the\count0}}}\next}
}
%
%

%
%
%
%

\gdef\prlmode{F}
\long\def\optional#1{}
%
%
\let\narrowequiv=\equiv
\def\equiv{\;\narrowequiv\;}

\def\dag{\dagger}     
\fontdimen16\tensy=2.7pt\fontdimen17\tensy=2.7pt 



%

%
%

%
%
%
\def\boxit#1#2{
        $$\vcenter{\vbox{\hrule\hbox{\vrule\kern3pt\vbox{\kern3pt
        \hbox to #1truein{\hsize=#1truein\vbox{#2}}\kern3pt}\kern3pt\vrule}
        \hrule}}$$
}




%




\def\splitexact#1#2{\mathrel{\mathop{\null{
\lower4pt\hbox{$\rightarrow$}\atop\raise4pt\hbox{$\leftarrow$}}}\limits
^{#1}_{#2}}}

%
%

%
%
%
%

\def\dd{\mskip 1.3mu{\rm d}\mskip .7mu} 



%
%

\def\IM{isomorphism}

%
%

\ifx\boringfonts\fonttest
\font\blackboard=cmssbx10 \font\blackboards=cmssbx10 at 7pt  
\font\blackboardss=cmssbx10 at 5pt
\else
\font\blackboard=msym10 \font\blackboards=msym7   
\font\blackboardss=msym5
\fi
\newfam\black
\textfont\black=\blackboard
\scriptfont\black=\blackboards
\scriptscriptfont\black=\blackboardss


%
\ifx\boringfonts\fonttest
\font\gothic=cmssbx10 \font\gothics=cmssbx10 at 7pt  
\font\gothicss=cmssbx10 at 5pt
\else
\font\gothic=eufm10 \font\gothics=eufm7
\font\gothicss=eufm5
\fi
\newfam\gothi
\textfont\gothi=\gothic
\scriptfont\gothi=\gothics
\scriptscriptfont\gothi=\gothicss

{\catcode`\@=11\gdef\oldcal{\fam\tw@}}
\newfam\curly
\ifx\boringfonts\fonttest\else
\font\curlyfont=eusm10 \font\curlyfonts=eusm7
\font\curlyfontss=eusm5
\textfont\curly=\curlyfont
\scriptfont\curly=\curlyfonts
\scriptscriptfont\curly=\curlyfontss
\def\cal{\fam\curly\relax}
\fi
%

\ifx\boringfonts\fonttest\else\fi

\global\newcount\pnfigno \global\pnfigno=1
\newwrite\ffile
\def\pfig#1#2{Fig.~\the\pnfigno\pnfig#1{#2}}
\def\pnfig#1#2{\xdef#1{Fig. \the\pnfigno}%
\ifnum\pnfigno=1\immediate\openout\ffile=figs.tmp\fi%
\immediate\write\ffile{\noexpand\item{\noexpand#1\ }#2}%
\global\advance\pnfigno by1}

%
%
\def\tfig#1{Fig.~\the\pnfigno\xdef#1{Fig.~\the\pnfigno}\global\advance\pnfigno
by1}

%
%
%
%
\def\figI{y}
\def\ifigure#1#2#3#4{
\midinsert
\ifx\figflag\figI
 \ifx\htflag\figI
 \vbox{
  \href{file:#3}
{Click here for enlarged figure.}}
 \fi
 \vbox to #4truein{
 \vfil\centerline{\epsfysize=#4truein\epsfbox{#3}}}
\else
\vbox to .2truein{}
\fi
\narrower\narrower\noindent{\bf #1:} #2
\endinsert
}








%
%

%


\def\inbar{\,\vrule height1.5ex width.4pt depth0pt}
\def\IB{\relax{\rm I\kern-.18em B}}
\def\IC{\relax\hbox{$\inbar\kern-.3em{\rm C}$}}
\def\ID{\relax{\rm I\kern-.18em D}}
\def\IE{\relax{\rm I\kern-.18em E}}
\def\IF{\relax{\rm I\kern-.18em F}}
\def\IG{\relax\hbox{$\inbar\kern-.3em{\rm G}$}}
\def\IH{\relax{\rm I\kern-.18em H}}
\def\II{\relax{\rm I\kern-.18em I}}
\def\IK{\relax{\rm I\kern-.18em K}}
\def\IL{\relax{\rm I\kern-.18em L}}
\def\IM{\relax{\rm I\kern-.18em M}}
\def\IN{\relax{\rm I\kern-.18em N}}
\def\IO{\relax\hbox{$\inbar\kern-.3em{\rm O}$}}
\def\IP{\relax{\rm I\kern-.18em P}}
\def\IQ{\relax\hbox{$\inbar\kern-.3em{\rm Q}$}}
\def\IR{\relax{\rm I\kern-.18em R}}
\font\cmss=cmss10 \font\cmsss=cmss10 at 10truept
\def\IZ{\relax\ifmmode\mathchoice
{\hbox{\cmss Z\kern-.4em Z}}{\hbox{\cmss Z\kern-.4em Z}}
{\lower.9pt\hbox{\cmsss Z\kern-.36em Z}}
{\lower1.2pt\hbox{\cmsss Z\kern-.36em Z}}\else{\cmss Z\kern-.4em Z}\fi}
\def\IGa{\relax\hbox{${\rm I}\kern-.18em\Gamma$}}
\def\IPi{\relax\hbox{${\rm I}\kern-.18em\Pi$}}
\def\ITh{\relax\hbox{$\inbar\kern-.3em\Theta$}}
\def\IOm{\relax\hbox{$\inbar\kern-3.00pt\Omega$}}


\long\def\optional#1{}
\def\pagin#1{}





\lref\CrHo{M. C. Cross and P. C. Hohenberg, {\sl Rev. Mod. Phys.} {\bf 65},
(1993).}

\lref\green{The kernels in \interface\ are
$$J_{ij}({\bf x},{\bf y})={\delta_{ij}\over|{\bf x}-{\bf y}|}+{
({\bf x}-{\bf y})_i({\bf x}-{\bf y})_j\over|{\bf x}-{\bf y}|^3},$$
$$K_{ijk}({\bf x},{\bf y})=-6{({\bf x}-{\bf y})_i({\bf x}-{\bf y})_j
({\bf x}-{\bf y})_k\over|{\bf x}-{\bf y}|^5}.$$
}

\lref\Chandra{S. Chandrasekhar, {\sl Hydrodynamic and Hydromagnetic
Stability} (Oxford University Press, Oxford, 1961).}

\lref\Plateau{ J. Plateau, {\sl Statique experimentale
et theorique des liquides soumis aux seules forces moleculaires}
(Gautier-Villars, Paris, 1873); 
Lord Rayleigh, {\sl Proc. Lond. Math. Soc.} {\bf 10}, 4 (1879); 
{\sl Phil. Mag.} {\bf 34}, 145 (1892).}

\lref\numrec{The front parameters were determined through a nonlinear
least-squares fitting algorithm [W.H. Press, B.P. Flannery, S.A.
Teukolsky, and W.T. Vetterling, {\sl Numerical Recipes in C}
(Cambridge University Press, New York, 1988)].}

\lref\BarZiv{ R. Bar-Ziv and E. Moses, {\sl Phys. Rev. Lett.} {\bf 73}, 1392 
(1994); R. Bar-Ziv, T. Tlusty,  and E. Moses, {\sl Phys. Rev. Lett.} {\bf 79},
1159 (1997).}

\lref\Mather{ P.~T. Mather, K.~P. Chaffee, T.~S. Haddad, and J.~D. Lichtenhan,
{\sl Polym. Preprints (Am. Chem. Soc., Div. Poly. Chem.)} 
{\bf 37}, 765 (1996).}

\lref\MT{ E.~S. Matsuo and T. Tanaka, {\sl Nature} {\bf 358}, 482 (1992); 
B. Barriere, K. Sekimoto, and L. Leibler, preprint 1996.} 

\lref\VE{ M. Renardy, {\sl J. Non-Newtonian Fluid Mech.} {\bf 59,} 267
(1995) and references therein.}

\lref\expts{ See for example W-K. Lee, K-L. Yu, and R.~W. Flumerfelt,
{\sl Int. J. Multiphase Flow} {\bf 7}, 385 1981.}

\lref\PNS{ P. Nelson, T. Powers, and U. Seifert, 
{\sl Phys. Rev. Lett.} {\bf 74}, 3384 (1995), cond-mat/9410104;
R. Granek and Z. Olami, 
{\sl J. Phys. II France} {\bf 5}, 1348 (1995);
R.~E. Goldstein, P. Nelson, T.~R. Powers, and U. Seifert, 
{\sl J. Phys. II France} {\bf 6}, 767 (1996), cond-mat/9510093.}

\lref\SBL{ H.~A. Stone, B.~J. Bentley, and L.~G. Leal, 
{\sl J. Fluid Mech.} {\bf 173,}
131 (1986).  See also the seminal work of G.~I. Taylor, {\sl Proc. Roy. Soc.
A} {\bf 146}, 501 (1934).}

\lref\DeeL{ G. Dee and J.~S. Langer, 
{\sl Phys. Rev. Lett.} {\bf  50}, 383 (1983); E. Ben-Jacob, H.
Brand, G. Dee, L. Kramer, and J.~S. Langer, 
{\sl Physica} {\bf 14D}, 348 (1985); W. van Saarloos, {\sl Phys. Rev. A}
{\bf 37}, 211 (1988); {\bf 39}, 6367 (1989).}

\lref\Saville{ S. Sankaran and D.~A. Saville, {\sl Phys. Fluids A} {\bf 5},
1081 (1993).}

\lref\Fish{ R. Fisher, {\sl Ann. Eugenics} {\bf 7}, 355 (1937);
A. Kolmogorov, I. Petrovsky, and N. Piskunov, 
{\sl Bull. Univ. Moskou, Ser. Internat., Sec. A} {\bf 1}, 1 (1937).}

\lref\Goldenfeld{ F. Liu and N. Goldenfeld, {\sl Phys. Rev. A} {\bf 39},
4805 (1988).}

\lref\Tomo{ S. Tomotika, {\sl Proc. Roy. Soc. Lond.} {\bf A150}, 322 (1932).}

\lref\Lamb{ H. Lamb, {\sl Hydrodynamics}, 6th ed.,
(Cambridge, Cambridge University Press, 1993).}

\lref\Limat{ L. Limat, P. Jenffer, B. Dagens, E. Touron, M. Fermigier,
and J.E. Wesfreid, {\sl Physica D} {\bf 61}, 166 (1992).}

\lref\Steen{ M.~J. Russo and P.~H. Steen, {\sl Phys. Fluids A} {\bf 1},
1926 (1989).}

\lref\drain{ S.~J. VanHook, M.~F. Schatz, W.~D. McCormick, J.~B. Swift,
and H. Swinney, {\sl Phys. Rev. Lett.} {\bf 75}, 4397 (1995).}

\lref\typeI{S.~J. Di Bartolo and A.~T. Dorsey, {\sl Phys. Rev. Lett.} 
{\bf 77,} 4442 (1996), cond-mat/9607023.}

\lref\pg{T.~R. Powers and R.~E. Goldstein, {\sl Phys. Rev. Lett.}
{\bf 78,} 2555 (1997), cond-mat/9612169.}

\lref\memdyn{For other applications and tests of theories of membrane
dynamics see {\it e.g.} U. Seifert, {\sl Adv. in Phys.} {\bf 46}, 13 
(1997) and M.~A. Peterson, {\sl Phys. Rev. E} {\bf 53}, 731 (1996).}

\lref\stonel{H.~A. Stone and L.~G. Leal, {\sl J. Fluid Mech.}
{\bf 198}, 399 (1989).}

\lref\pozrikidis{C. Pozrikidis, 
{\sl Boundary Integral and Singularity Methods for
Linearized Viscous Flow} (Cambridge University Press, Cambridge, 1992).
See also A.~Z. Zinchenko, M.~A. Rother, and R.~H. Davis, {\sl Phys. Fluids}
{\bf 9}, 1493 (1997).}

\lref\tanzosh{J. Tanzosh, M. Manga and H.A. Stone, 
{\sl Proceedings of Boundary Element Technologies
VII}, C.A. Brebbia and M.S. Ingber, Eds. (Computational Mechanics
Publications, Boston, 1992), p. 19.} 

\lref\zhangs{D.~F. Zhang and H.~A. Stone, {\sl Phys. Fluids} {\bf 9},  (1997).}

\lref\lee{S.~H. Lee and L.~G. Leal, {\sl J. Colloid. Interface Sci.} 
{\bf 87}, 81 (1982).}

\lref\eggers{J. Eggers and T.~F. DuPont, {\sl J. Fluid Mech.} {\bf 262}, 
205 (1994); D.~T. Papageorgiou, {\sl J. Fluid Mech.} {\bf 301}, 109 (1995).}

\lref\satellite{Many of the breakup events involved smaller, satellite
droplets, the analysis of which we have not pursued.}

\lref\lowvals{It is well known that numerical boundary integral methods
become less accurate at small $\lambda$ \pozrikidis.  
We found that for $\lambda=0.001$ the volume of the drop changed by 
about $10\%$ during the continuous evolution. 
}

\lref\tahadji{ Fig. 3c of M. Tjahjadi, H. A. Stone, and J. M. Ottino, 
{\sl AIChE J.} {\bf 40}, 385 (1994).}

\lref\plasma{R.~J. Briggs, {\sl Electron Stream Interactions with Plasmas,}
(MIT Press, Cambridge, MA 1964); E.~M. Lifshitz and L.~P. Pitaevskii, 
{\sl Physical Kinetics} (Butterworth-Heinemann Ltd., Oxford 1981), 
sec. 62--63. }

\lref\asymp{There is another possible source of error.  In the MSC picture,
the front accelerates to the MSC velocity with a relaxation time
that scales as $q''^{-1}$ \DeeL\ \plasma.  Thus, 
when retraction is unimportant, we would expect our numerical
results to lie below the MSC curve.  This discrepancy should
be greatest for the smallest values of $\lambda$ we considered, since it
is precisely those values for which $q''$ is getting small \pg.  However,
we saw no evidence of acceleration of the front in the numerical 
calculations.}

\lref\AW{D.~G. Aronson and H.~F. Weinberger, in {\sl Partial Differential
Equations and Related Topics,} J.~A. Goldstein, Ed. (Springer-Verlag, 
Heidelberg) 5 (1975).}

\def\testp{T}

\Title{ }{\vbox{\centerline{Propagation of a Topological Transition:}
\vskip2pt\centerline{the Rayleigh Instability}
}}

\centerline{\bf Thomas R. Powers$^{1}$, 
Dengfu Zhang$^{2,*}$,  
Raymond E. Goldstein$^{1,3,\dag}$, and
Howard A. Stone$^{2}$}\smallskip
\centerline{\sl $^1$Department of Physics, 
University of Arizona, Tucson, AZ  85721;}
\centerline{\sl $^{2}$Division of Engineering and Applied Sciences, 
Harvard University, Cambridge, MA 02138;}
\centerline{\sl $^3$Program in Applied Mathematics, University of Arizona,
Tucson, AZ 85721.}
\bigskip\bigskip

The Rayleigh capillary instability of a cylindrical interface between two 
immiscible fluids
is one of the most fundamental in fluid dynamics.  As Plateau 
observed from energetic considerations
and Rayleigh clarified through hydrodynamics, 
such an interface is
linearly unstable to fission due to surface tension.
In traditional descriptions of this instability it occurs
everywhere along the cylinder at once, triggered by
infinitesimal perturbations.  
Here we explore in detail a recently conjectured
alternate scenario for this instability: front propagation.
Using boundary integral techniques for Stokes flow, we provide
numerical evidence that the {\it viscous} Rayleigh instability 
can indeed spread
behind a front moving at constant velocity, in some cases leading to a 
periodic sequence of pinching events.
These basic results are in quantitative agreement with the marginal
stability criterion, yet there are important qualitative differences
associated with the discontinuous nature of droplet fission.
A number of experiments immediately suggest themselves in light of 
these results.

\vskip 1cm

\noindent{$^*$} Present address:  Adapco, 60 Broadhollow Rd., 
Melville, NY 11747.

\ifx\prlmode\testp
\noindent {\sl PACS: 

}\fi
\ifx\answ\bigans \else\noblackbox\fi
\Date{\today }\noblackbox
\noblackbox

\hfuzz=3truept
\def\pagin#1{}

Recently, Bar-Ziv and Moses discovered the ``pearling instability''
of lipid bilayer membrane tubes, in which the application of laser
tweezers to the membrane induces a periodic modulation in the radius \BarZiv.  
Theoretical explanations \PNS\ for this behavior invoke a 
tension created in the membrane by the laser tweezers.
Were this an interface between two immiscible fluids, 
such tension would induce capillary breakup of the cylinder into
droplets {\it via} the Rayleigh instability \Plateau. 
In the pearling phenomenon however,
breakup is prevented by the membrane bending elasticity.
A more striking difference is that the pearling instability propagates---the 
modulated state is observed to
invade the uniform cylindrical region at a constant velocity.
This propagation is totally unlike traditional descriptions of the 
Rayleigh instability \Chandra, in which perturbations grow uniformly along the 
length of the tube.  

In a previous publication \pg, we conjectured
that the viscous Rayleigh instability can propagate as well.
Propagating fronts in which a new stable state invades an unstable region
are a common feature of overdamped systems \CrHo, arising
in such diverse situations as reaction-diffusion systems, 
dendritic growth, and, more recently, type-I superconductors \typeI.    
These previously well-studied
examples all display a 
{\it continuous} evolution, in contrast to the discontinuous evolution
associated with drop fission.  
Below we show by numerical methods that despite this fundamental difference,
the Rayleigh instability of a cylindrical interface can indeed propagate.
Furthermore, the coarse features of this behavior can be described by 
the {\it marginal stability criterion} (MSC), an analytical method which 
has found wide applicability in the simpler situations mentioned above \DeeL.

The discontinuous evolution associated with rupture of the thread
can have significant implications for the possible existence of a
propagating front.
As a droplet pinches off from the main body, the two tips of the
broken neck recede from the pinching point; 
if the retracting end overtakes the front, propagation will be spoiled.
Experiments on the breakup and relaxation of elongated drops suspended in
an outer fluid
\SBL\ reveal that this competition depends on the relative viscosity 
of the two fluids.  When the drop viscosity $\eta^-$ 
is much smaller than the outer viscosity $\eta^+,$ the drop breaks
before its ends have time to retract; in the other extreme, the
ends retract significantly before breaking off (for long enough drops).  
We shall see
below that when the viscosity ratio $\lambda=\eta^-/\eta^+$ 
is such that retraction
is slow on the time scale for breakup, there is a propagating front 
moving with constant velocity (\tfig\snapshots).
No analytical solution is known for the complex shape evolution
shown in \snapshots.  Indeed, this is the case in most examples of 
front propagation, where, however, the MSC nearly always provides correct
predictions for front speed.  In our studies of drop breakup 
below we find that the front speed is rather accurately
given by the linear MSC, a remarkable result in light of the strongly
nonlinear, singular shape evolution behind the front.
In addition to the front velocity, we compute the time between primary 
pinching events in the breakup of the cylinder using the MSC quantities 
and find good agreement with the numerical calculations. 

\ifigure\snapshots{Sequence of drop shapes for viscosity 
ratio $\lambda=0.05$ at $t_n=6.67n\eta^+R/\gamma$;
$n=1,2,3,..., 15$ from top to bottom.
To illustrate the complete evolution, we have drawn the 
daughter droplets (but not the satellite droplets).  However,
the evolution of each connected component was computed independently
of the others.}
{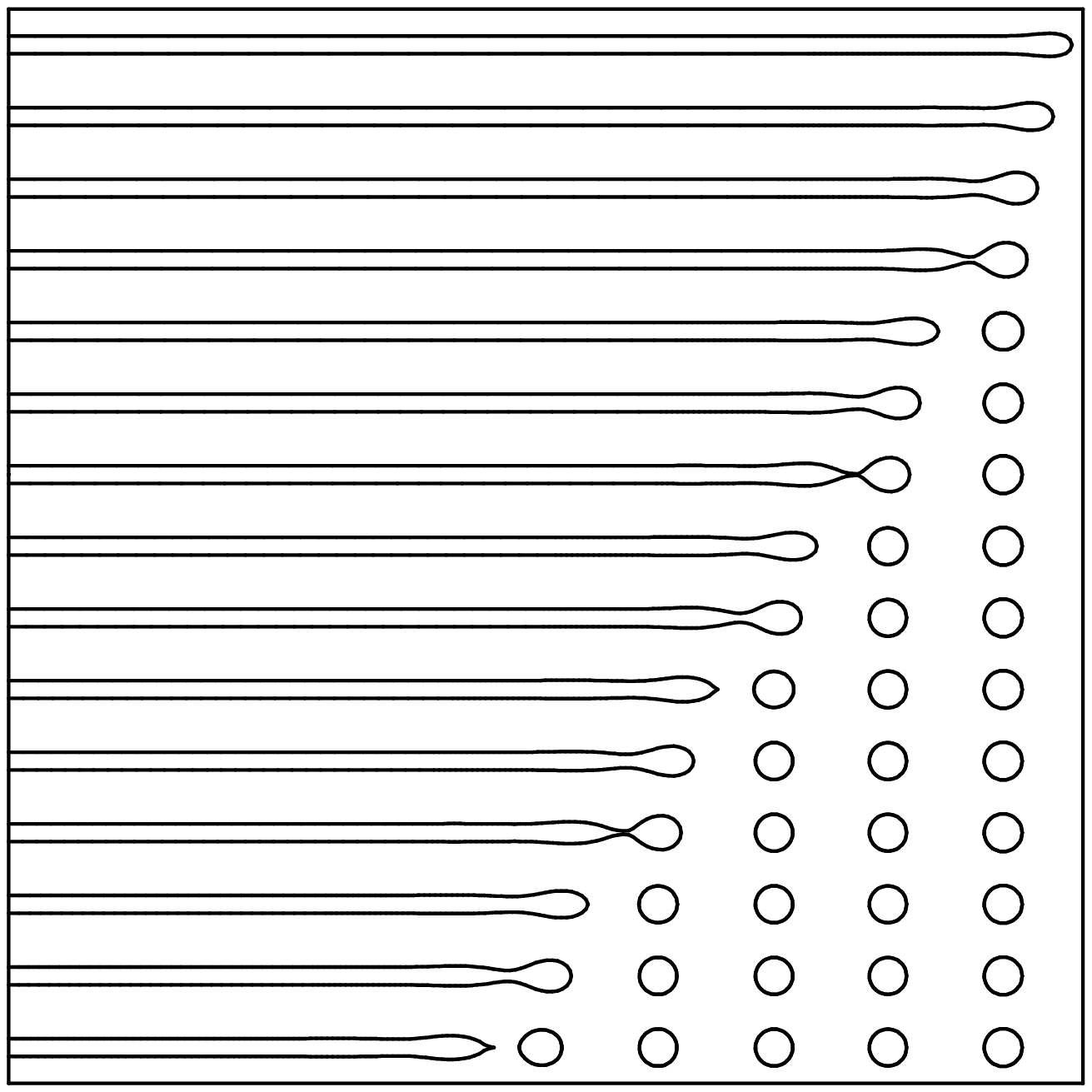}{6.0}

{\bf Drop evolution.}  We look for the 
propagation of the Rayleigh instability in a numerical simulation of a 
long, axisymmetric, approximately cylindrical drop, initially 
stationary and with hemispherical caps on the ends.  
The axis
of the cylinder defines the $x$-direction, so that the shape is
given by the surface of revolution generated by the radius $r(x)$ 
We work in the overdamped limit in which the inertial terms of the
Navier-Stokes equations may be disregarded.
Therefore, the outer and inner 
fluids with velocities ${\bf u}^\pm$ and pressures $p^\pm$
are described by the Stokes equations and the constraint 
of incompressibility
\eqn\stokes{\eta^{\pm}\nabla^2 {\bf u}^\pm = {\bf \nabla} p^\pm, \quad 
{\bf \nabla}\cdot {\bf u}^\pm=0.}
Propagation in the inertial regime is complicated by the presence of
dispersive capillary waves, and we leave this case and the case 
of net flow (like a jet) for future work. 
The boundary conditions at the interface $S$ are continuity of fluid velocity
${\bf u}^+|_S={\bf u}^-|_S$,
continuity of tangential
stress, and the jump in normal stress:
\eqn\stress{n_i(\sigma^+_{ij}-\sigma^-_{ij})|_S=-2\gamma H n_j,}
where ${\bf n}$ is the outward surface normal, $H$ is the mean curvature,
$\gamma$ is the (constant) interfacial tension,
and the stress tensors are
$\sigma^\pm_{ij}=
\eta^\pm(\nabla_i u^\pm_j + \nabla_j u^\pm_i) -p^\pm \delta_{ij}.$   
For an axisymmetric shape with radius $r(x)$, and with
$r_x\equiv \partial r/\partial x$, the mean curvature is
$H={(r_{xx}/2) 
{(1+r_x^2)^{-3/2}}}-{{(2r)^{-1}(1+r_x^2)^{-1/2}}}.$ 
The only important material
parameters are therefore the surface tension $\gamma$ and the viscosities
$\eta^+$ and $\eta^-.$

Note that \stress\ gives rise to the {\it absolute} (rather than convective) 
instability of a stationary cylindrical interface.
When the fluid velocity is zero, 
the pressure jump across the interface is $\Delta  p=2\gamma H$.  Consider an
axisymmetric perturbation in which the radius is
slightly pinched to form a neck.   If the disturbance is of sufficiently 
long wavelength, the magnitude of the curvature is increased,
thus raising the pressure in the neck. Fluid is therefore forced out of 
the pinched region, leading to growth of the perturbation.

The challenging numerical task of solving the three-dimensional 
Stokes equations is greatly simplified by the 
boundary integral technique, in which \stokes\ and \stress\ are recast
as an integral equation for quantities on the interface \pozrikidis.
This approach removes one spatial dimension from the problem, leading to 
\eqnn\interface
$$\eqalignno{{(1+\lambda)\over2} u_j ({\bf y})  &= 
{ \gamma \over 4\pi\eta^+}\int_{S} H({\bf x}) n_i({\bf x}) J_{ij}(
{\bf x},{\bf y})  \dd S({\bf x})\cr &+ 
{(1-\lambda)\over 4\pi}{\cal P.V.} 
\int_{S}  u_i({\bf x}) n_k({\bf x})  K_{ijk}({\bf x},{\bf y}) \dd 
S({\bf x}),&\interface}$$
where ${\bf y}$ is a point on the interface $S$, ${\bf u}({\bf y})$ is
the velocity of the interface at ${\bf y}$, ${\cal P.V.}$ denotes the 
principal value, and the kernels $J_{ij}$ and $K_{ijk}$ are
Green's functions \green.

Axisymmetry reduces our problem to an integral equation
with one space and one time dimension. This equation is solved numerically 
by standard adaptive-grid techniques \tanzosh\ to yield the interfacial
velocities, and so the drop shape, as a function of time.  
This procedure can describe the continuous
motion of the interface only; to describe the breakup of the interface
we simply demand that pinching occurs whenever the radius $r(x)\leq 0.005 R$, 
where $R$ is the initial cylinder radius.  Although somewhat arbitrary,
we expect this choice for the cutoff leads to little error in the gross
evolution since further decreases in the neck 
radius occur rapidly owing to the large velocity and curvature gradients 
near the pinch point. Furthermore, the size of the region in which 
these quantities grow large is small 
\eggers.  In any case,
these same factors make it difficult to do numerical calculations when the 
neck approaches rupture.  Once a droplet has pinched off, we neglect it 
in further 
calculations, since experience has shown the the evolution of the droplet 
that pinches off has little effect on the evolution of the main drop \stonel.

Our findings for the shape evolution are in qualitative accord with 
previous investigations
\SBL.  At higher values of $\lambda,$ {\it i.e.} around $\lambda=10,$
the evolution is dominated by retraction.  As an example, \tfig\results\
shows that the $\lambda=10$ drop simply retracts and shows no 
sign of developing a neck 
in the time it takes the $\lambda=0.1$ drop to break a few times.
\tfig\dropend\ shows a magnified view of the evolution of these droplets.
Note that at a given time the radius $r(x)$ of the $\lambda=10$ drop 
grows monotonically in $x$
to its maximum value, but the radius for $\lambda=0.1$ is modulated;
these undulations are the seeds of the capillary instability.

When the outer viscosity is 
not too small, we can estimate the retraction speed by assuming the 
bulge on the drop end is spherical
and balancing tension with drag \pg.  This leads to a velocity 
that scales like $\gamma/\eta^+$, since the drag is controlled by 
the outer viscosity \Lamb.  
This estimate only accounts for the dissipation due to 
the flows in the two fluids
induced by dragging a spherical interface through the outer fluid; 
it disregards the contribution due to shape change, which become 
comparable when $\lambda\ge1$.
Indeed, when there is no outer fluid all
the dissipation takes place inside the retracting end.
The estimate is borne out by our simulations:
if we take the retraction velocity to be the speed of the maximum
of $r(x)$ nearest the end of the drop (\tfig\front), 
we find retraction velocities
that range from $0.2\gamma/\eta^+$ at $\lambda=0.005$ to $0.1\gamma/\eta^+$ for
$\lambda=10.$

\ifigure\results{Numerical results.  Propagation of the Rayleigh instability
for $\lambda=0.1$ (left).  Retraction-dominated dynamics for $\lambda=10.$
(right).  $t_n=6.67n\eta^+R/\gamma,$
where $n=1,2,3,..., 15$ labels the shapes from the top.}{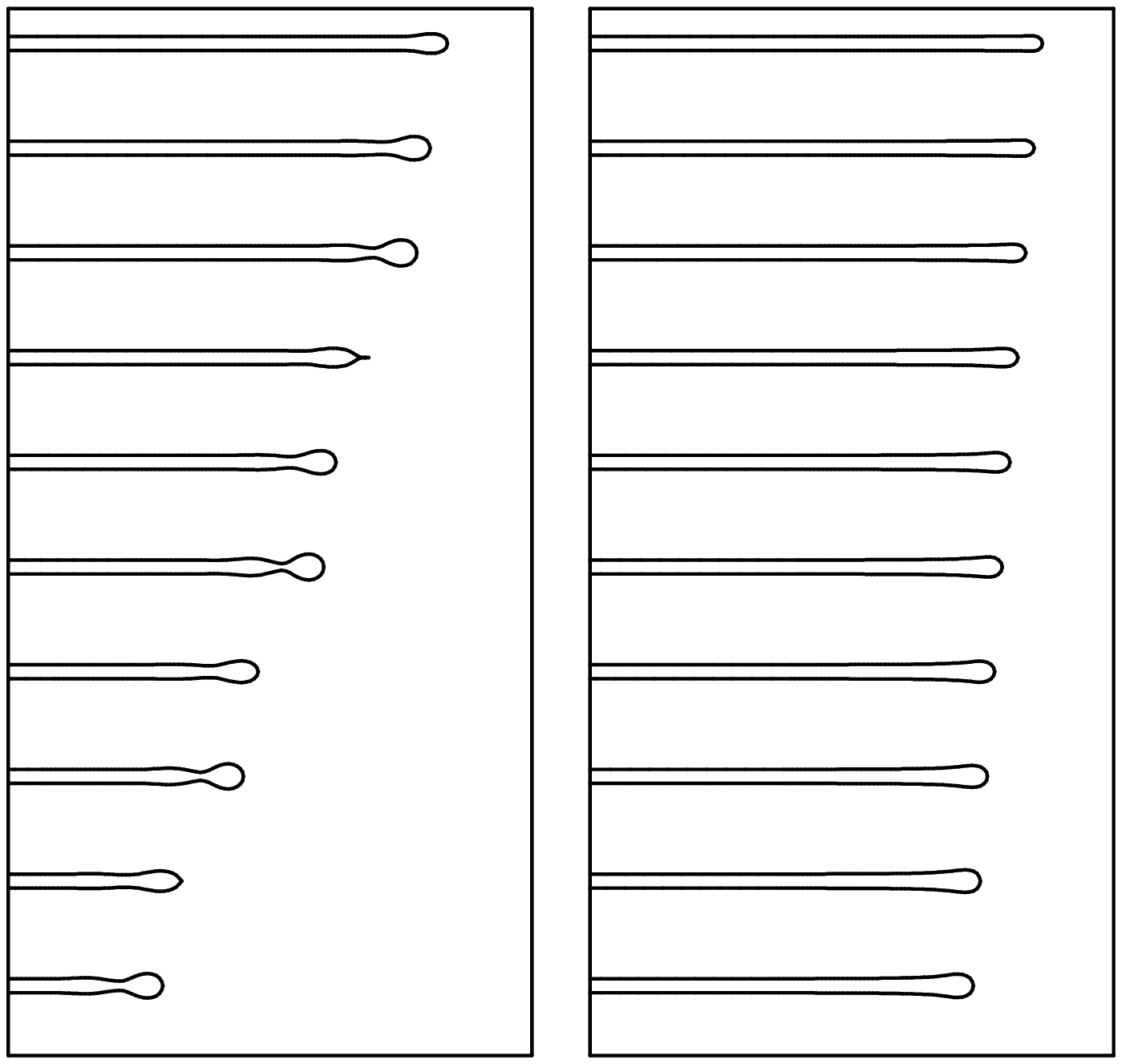}{6.0}

The front region connects the uniform cylinder, which has not 
yet undergone the Rayleigh instability, to the growing bulge, which
eventually pinches off (\front).  We extract the front from the 
sequence of drop shapes by choosing for each shape a window with one
edge such that $r(x)$ is within numerical accuracy of the unperturbed radius
$R$, and the other edge with $|r(x)-R|/R$ reaching a small value
$a^*$ (typically $0.1$--$0.15$).  
The drop radius in this window
is fit \numrec\ with an exponential envelope 
times a sinusoidal function of the form
$r(x)=a_1+a_2\exp(-q''x)\cos(q'x-a_3).$
We find that the fit parameters $a_1, a_2, a_3, q', q''$ 
do not change with small changes
in the choice of fitting window or small changes in the initial shape
of the ends of the drop.   
The front speed is the velocity of the point $x_f(t)$ for
which the envelope function $a_2\exp(-q''x_f(t))=a^*$.

\ifigure\dropend{Enlarged view of the droplet ends of \results\ 
for $\lambda=0.1$ and
$\lambda=10.$  $t_n=(37.5+3n)\eta^+R/\gamma,$ where $n=1,2,...,6$ labels 
the shapes.}{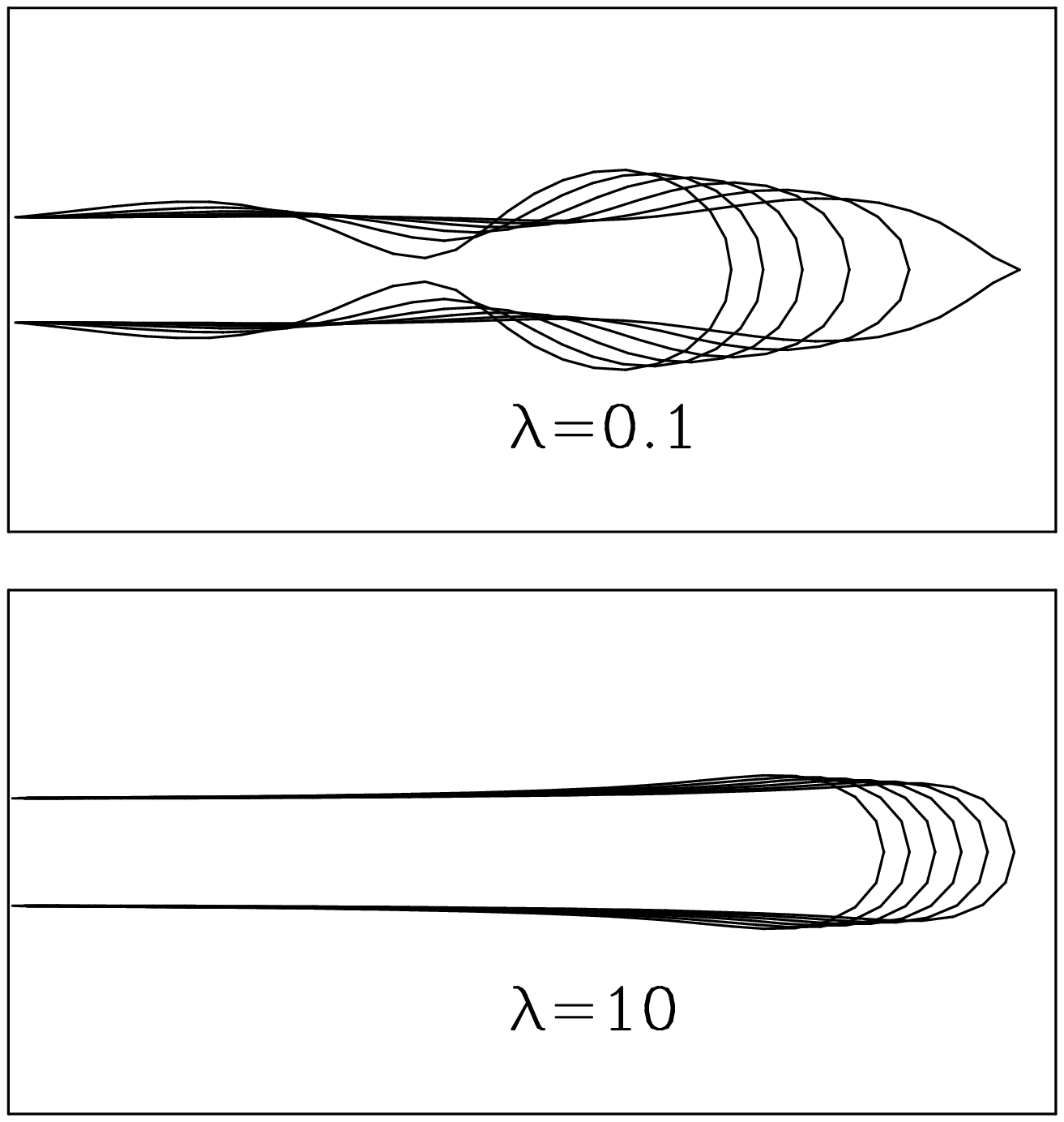}{6.0}

Turning now to the analysis of the fronts,
we find a smoothly moving front for a range of
viscosity ratios, from $\lambda=0.005$ to $\lambda=1.0.$  For example,
the top graph
in \tfig\mscquant\ shows the front position as a function of time for 
$\lambda=0.05.$
At low values of $\lambda,$ around $\lambda=0.005$ \lowvals, the ends do not
retract much before the drop pinches (\results).  This is the 
``end-pinching'' behavior described in \SBL:  the daughter 
droplets pinch off one at a time in a periodic fashion at the ends of the 
main drop \satellite.
Despite the discrete nature of these
events, there is a front moving smoothly at constant velocity.  We therefore
interpret this end-pinching at low viscosity ratio
as a front of the Rayleigh instability.

\ifigure\front{A snapshot of the end of the main drop for $\lambda=0.05,$
in which the aspect ratio has been distorted to show the front more clearly. 
The center of the drop is at $x=0.$
The darkened line is the front region; the inset is a magnified view of
this region (solid line) and a fit to the front (dotted line).  
The retraction speed is taken to be the
velocity of the maximum marked with a plus sign.}{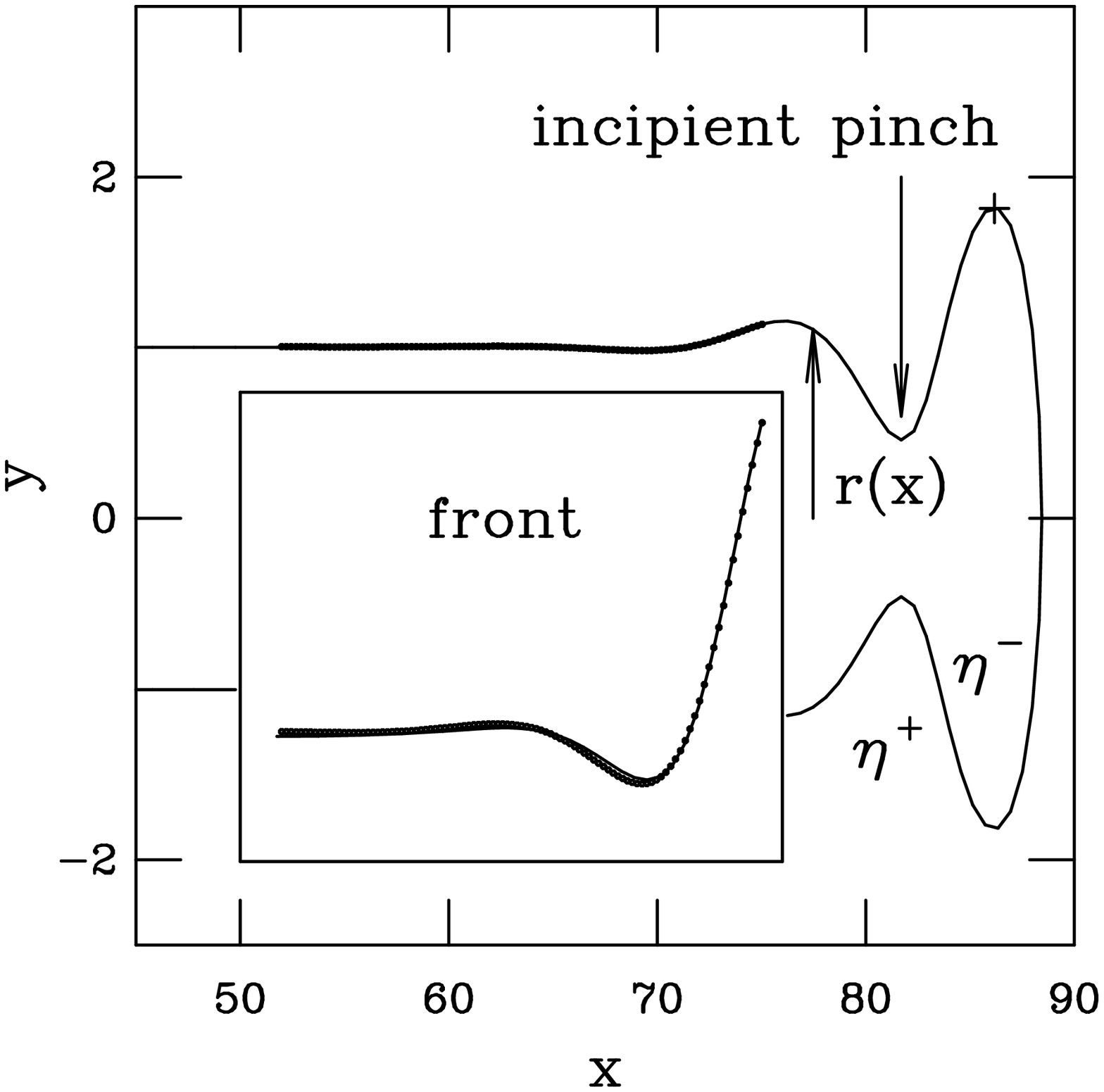}{6.0}

As $\lambda$ increases, retraction
becomes more apparent in the time it takes the droplets to break off.  
When $\lambda$ reaches a value at which the retraction speed becomes comparable
to the speed at which the pinches are spreading (around $\lambda=1$), 
we find that the drop evolution can not be described as a smoothly
moving front.  The details of this process are complicated.
For $\lambda\simeq 0.2$, the periodic
behavior of the breakup ceases.  The droplets that pinch off are not of
uniform size and do not pinch off at a constant rate.
Nevertheless, we are still able to fit the advancing profile to a front
moving smoothly with constant velocity.
When $\lambda>1.0,$ this smoothly propagating front behavior is lost.
Around this value of $\lambda$ we also observe 
necks that start to form at one point but then heal, leading 
to breakup at a different point along the drop.  
This behavior is reminiscent of that
seen in the experiments of Tjahjadi {\it et al.}, who observed
that a drop stretched
to a cylindrical shape just beyond the critical aspect ratio for breakup does
{\it not} break \tahadji. Instead, two necks
form and then disappear, and the elongated drop retracts to a sphere
without breaking.  We can offer no explanation of this behavior, and merely
note that when retraction is faster than the pinching process,
there is no reason to expect the simple picture of the Rayleigh instability 
outlined in the last section to apply.

{\bf Front velocity and the MSC.}  
For completeness, we review first the basic facts of
the marginal stability criterion.  In the leading edge of the front,
the amplitude $u(x,t)\equiv r(x,t)-R$ is small and can be represented
as a linear combination of Fourier modes each having the form
$u(x,t)\sim\exp(\omega(q^*)t+i q^* x)$, in which $\omega(q)$, possibly
complex, is the
linear growth rate, and the real and imaginary parts of 
the wavevector $q=q'+iq''$
describe the periodicity and sharpness of the mode.  
The velocity of the exponential envelope of
any mode is $v_q={\rm Re} \omega/{\rm Im} q$.  
The MSC selects a unique mode $q^*$ (and hence velocity $v^*$) 
by the condition that the envelopes of modes nearby in $q$ neither
outrun nor fall behind that of $q^*$.
This provides two additional relationships between $v$ and $q$, leading
to the conditions
\eqn\msc{v^*={{\rm Re} \omega^*\over {\rm Im} q^*},\ \ \ 
{\rm Im} {\partial\omega^*\over\partial q}=0,
\ \ \  v^*={\rm Re}{\partial\omega^*\over\partial q},}
where $\omega^*=\omega(q^*)$ \DeeL\plasma.
A rigorous proof of the validity of this approach is known only
for the Fisher-Kolmogorov equation \Fish,
$u_t=u_{xx}+u-u^3$, for which
it can be shown that sufficiently
localized initial conditions will evolve into a front moving at a 
unique speed $v=v^*$ \AW.  A very large number of partial differential
equations have been shown to be correctly described by this principle
or its nonlinear variants \DeeL, even when no 
analytic front solution is known.
There is, however, 
no general criterion to determine when the MSC is applicable, and thus
the question of its validity in this new setting 
of discontinuous evolution is of great interest.

The MSC quantities depend only on the growth rate $\omega$, which is known from
Tomotika's generalization \Tomo\ of Rayleigh's result to the two-fluid problem.
For a cylinder
of radius $R$, $\omega$ is a function of wavevector
$q$ and viscosity ratio $\lambda$ in the form 
$\omega(q,\lambda)=(\gamma/R\eta^+)\Lambda(q,\lambda)(1-q^2)$ .
The dynamical factor $\Lambda(q,\lambda)$, too lengthy to quote here, 
accounts for viscous dissipation, and the factor $\gamma(1-q^2)$ is 
associated with
the Young-Laplace force due to a constant-volume distortion of wave number $q$.
Inserting this growth rate into the MSC equations \msc\ yields
the front velocity previously derived in \pg\
as a function of viscosity ratio.  \tfig\velvslambda\ compares this
prediction with our present numerical calculations.  

Note that since the growth rate for the Rayleigh instability does not
contain the physical process of retraction, we expect the MSC velocity
to depart from the true front speed when the front and 
retraction speed are comparable.  This 
may be the source of disagreement between the MSC and the numerical 
simulations for $\lambda\ge0.5.$  As $\lambda$ is increased beyond this
value, retraction becomes more important until finally the simple picture
of a propagating Rayleigh instability becomes invalid.
At the lower end of the viscosity contrasts we studied, the boundary 
integral method starts to lose accuracy; since it is difficult to 
quantify these errors we cannot say if this is the source of disagreement
between the MSC and the numerical calculations \asymp.

\ifigure\mscquant{Top: results of boundary integral calculations and 
fitting routine for front position {\it vs.} time.
Bottom:  comparison of prediction of MSC and results of boundary
integral method for the initial front wavenumber $q'$ and front width
$q''.$ Both plots are for $\lambda=0.05.$  The simulation begins at
$t=0;$ by $t \gamma/(\eta^+R)=20$ transients from the initial shape
have died out.}{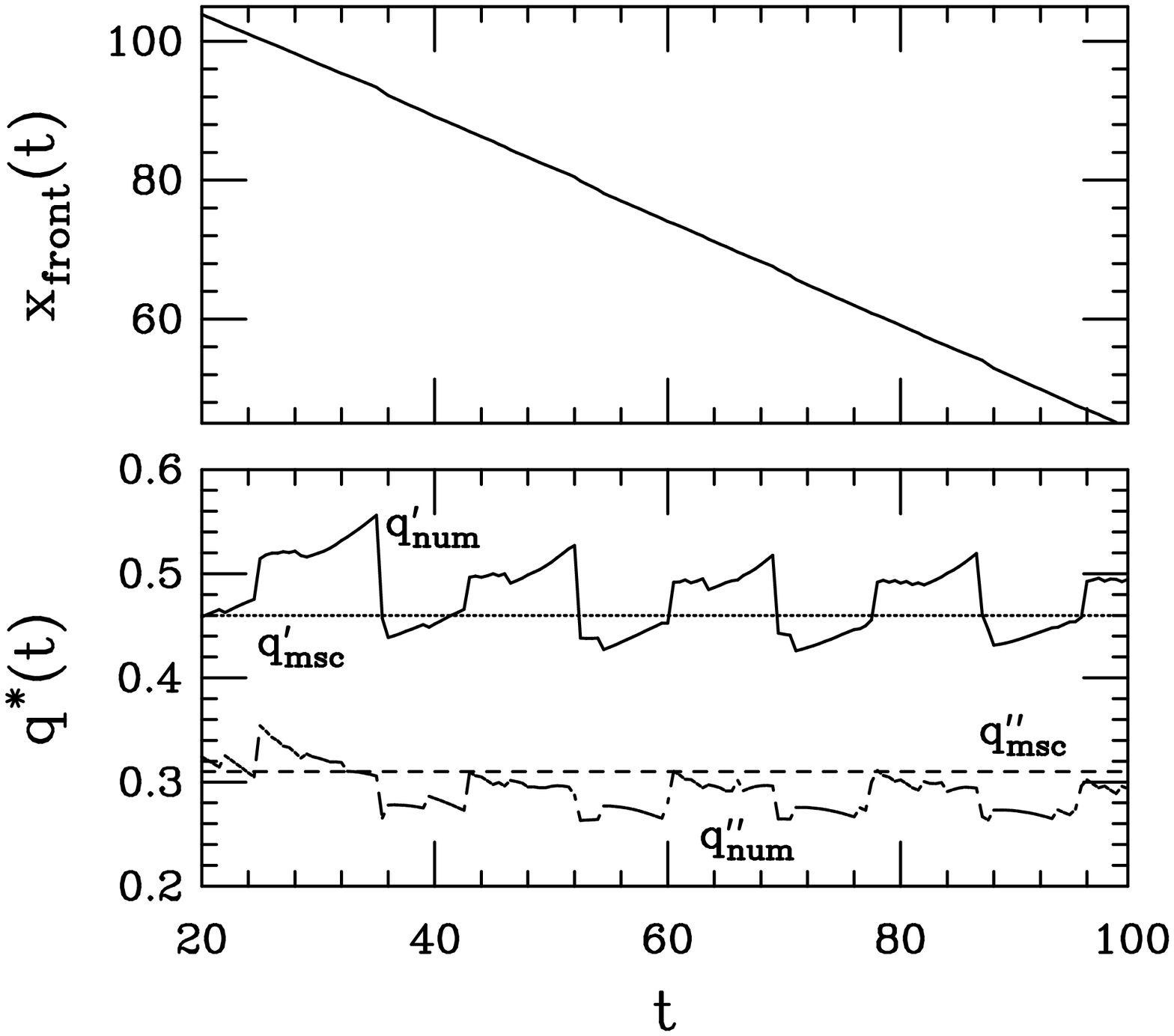}{6.0}

While the front position advances smoothly in accord with the MSC picture,
the behavior of the front shape encoded in the wavenumber $q'$
and the inverse front width $q''$ is more complicated.  We find that 
these quantities are not constant but rather oscillate with approximately 
the period of pinching; however, they are always near the MSC values.  
We show one example in the bottom graph of \mscquant.  We conclude that
the MSC captures
the gross features of the front shape but not the fine structure.

\ifigure\velvslambda{Front velocity $v$ vs. $\lambda.$  The circles are
the results of the numerical calculations, the line is the prediction of the
MSC from \pg.}{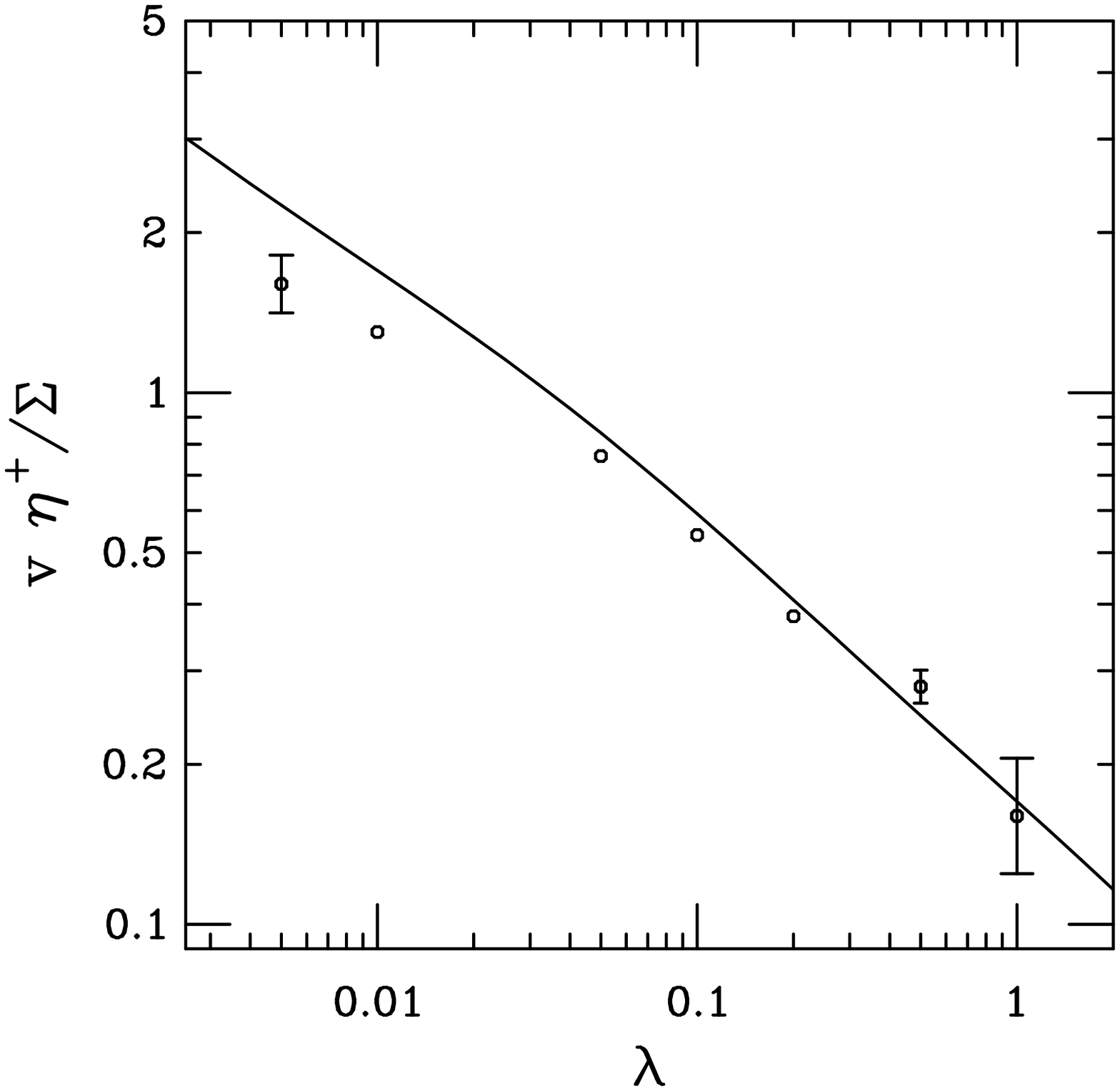}{6.0}

Given that we find the breakup process to be periodic in the 
viscosity range $\lambda=0.005$--$0.1,$ it is natural
to ask if the MSC can predict the time $t_{\rm pinch}$ 
between primary pinching events, 
once the transients due to the initial shape settle down.
A plausible but naive starting point is to use the shortest
characteristic growth time, 
$t_{\rm pinch}=\omega(q_{\rm max})^{-1},$
where $q_{\rm max}$ is the wavenumber of the fastest growing mode.
However, it is well known that the fastest growing mode is 
not the mode that is selected when the MSC is applicable.  In the 
standard MSC treatment, there are in fact {\it two} selected 
wavenumbers:  $q',$ the selected wavenumber in the {\it leading edge} of the 
front, and $q_0={\rm Im}(\omega(q^*)+i q^*v^*)/v^*$ \DeeL, 
the selected wavenumber in the {\it saturated} pattern.
Thus another candidate for the pinching time is 
$t_{\rm pinch}=2\pi /(q_0v^*).$  However, as we pointed out
in \pg, we do not expect $q_0$ to be relevant to the Rayleigh problem
since the saturated pattern has not continuously evolved from the leading edge
of the front.  Therefore
we expect the appropriate wavelength to be $2\pi/q'.$  In \tfig\pinchtimes\
we show that the numerical simulations clearly favor the relation 
$t_{\rm pinch}=2\pi/(q'v^*).$

\ifigure\pinchtimes{Various predictions for the primary pinch time.
Solid line is $2\pi/(q'v),$ dotted line is $2\pi/(q_0v),$ and the dashed
line is $1/\omega(q_{\rm max}).$  The circles are from the simulation.
The breakup is aperiodic for $\lambda\ge0.2,$ leading to a distribution 
of droplet sizes.}{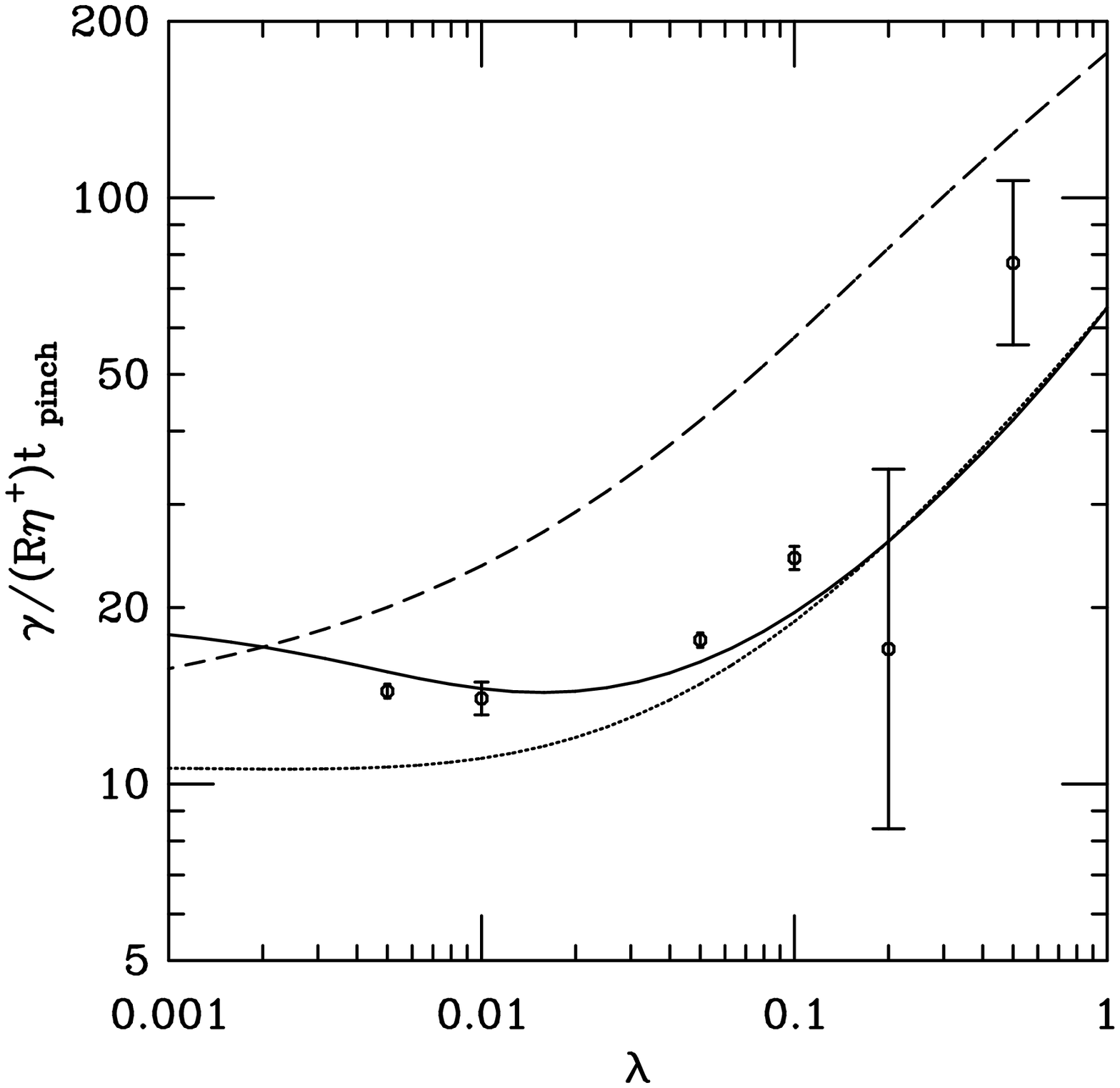}{6.0}

\suppress{There is another effect made more important by the breaking:
the possibility that transients do not die out in the time it takes
a droplet to form and break.
For example, one could imagine that the shape evolution 
asymptotically approaches a propagating front solution
but that each time a droplet breaks off the initial conditions are reset
before the propagating front has been reached.
In this situation the description of front propagation is inappropriate.
We shall see below that this is not the case, probably because 
the small region near the pinching neck does not have much effect 
on the rest of the shape.}
In conclusion, we have shown that the breakup of elongated drops
can proceed {\it via} front propagation for the viscosity ratios in the
range $\lambda=0.005$--$1.0.$  Remarkably, despite the complex nature
of the discontinuous dynamics, the marginal stability criterion gives
a good description of the gross quantitative features of the shape evolution,
such as the front velocity and shape, and the pinch-off time. 
There are many experimental systems which could be used to test our
results.  The most natural candidate is the four-roll mill, 
in which a planar hyperbolic flow stretches a spherical drop into a
long thread \SBL.  When this flow ceases, the relaxation or breakup
of the thread can be studied.  Other possibilities are capillary bridges
stabilized by electric or magnetic fields \Saville\ or very viscous jets.
Experimental confirmation of our picture of propagating topological
transitions would be an important
contribution to the understanding of interface motion in fluid dynamics.

\vfil
\eject
{\frenchspacing
\ifx\prlmode\testp\else\vskip1truein \leftline{\bf Acknowledgements}
\noindent\fi
TRP thanks Dartmouth College for hospitality.
This work was supported by NSF Presidential 
Faculty Fellowship grant DMR 93-50227 (REG),  
and the Petroleum Research Fund (Grant No. 28690-AC9) (DFZ and HAS). 
E-mail: powers@physics.arizona.edu, 
dengfu@adapco.com, gold@physics.arizona.edu, has@stokes.harvard.edu.

}

\listrefs

\bye